\documentclass[%
prb,
superscriptaddress,
reprint,
 amsmath,amssymb,
 aps,
]{revtex4-1}
\usepackage{graphicx}% Include figure files
\usepackage{dcolumn}% Align table columns on decimal point
\usepackage{bm}% bold math
\usepackage[mathlines]{lineno}% Enable numbering of text and display math
\usepackage{braket}
\usepackage[dvipsnames]{xcolor}
\usepackage{amsmath,amssymb,amsfonts} % Typical maths resource package
\usepackage{float}
\usepackage{bbm}
\usepackage{nicefrac}
\usepackage{wasysym}
\usepackage[normalem]{ulem}
\usepackage{hyperref}
\hypersetup{
	colorlinks=true,
	linkcolor=blue,
	filecolor=blue,
	urlcolor=blue,
	citecolor=blue,
}

\def\kk{{\bm k}}

\begin{document}

\title{Unified Multipole Bott Indices for Non-Hermitian Skin Effect in Different Orders}

\author{Wuping Yang}
\affiliation{School of Physics, Peking University, Beijing 100871, China}

\author{H. Huang}%
\email[Corresponding author: ]{huanghq07@pku.edu.cn}
\affiliation{School of Physics, Peking University, Beijing 100871, China}
\affiliation{Collaborative Innovation Center of Quantum Matter, Beijing 100084, China}
\affiliation{Center for High Energy Physics, Peking University, Beijing 100871, China}

\date{\today}% It is always \today, today,

\begin{abstract}
Non-Hermitian systems exhibit a distinctive phenomenon known as the non-Hermitian skin effect, where an extensive number of eigenstates become localized at the boundaries of a lattice with open boundaries. While the spectral winding number under periodic boundary conditions is a well-established topological indicator for predicting the skin effect in one-dimensional non-Hermitian systems, a suitable topological invariant to diagnose higher-order skin effects remains elusive. In this Letter, we propose a unified non-Hermitian multipole characterization framework that generalizes the concept of spectral winding to higher-order skin effects. Specifically, we develop a set of non-Hermitian multipole Bott indices capable of diagnosing skin effects of different orders. Our approach provides a comprehensive understanding of both first- and higher-order skin effects in non-Hermitian systems, offering new perspectives for exploring topological phenomena in non-Hermitian systems.
\end{abstract}

\maketitle

\textit{Introduction.}---Non-Hermitian (NH) Hamiltonians is an efficient framework to describe the coupling between materials and the environment \cite{RevModPhys.93.015005, doi:10.1080/00018732.2021.1876991,PhysRevLett.85.2478,PhysRevC.67.054322,Rotter_2009,PhysRevA.85.032111,Rotter_2015}, and has been experimentally implemented in a wide range of classical wave systems, including photonics \cite{RevModPhys.91.015006,feng2017Non, AosRnsurr2012Parity, doi:10.1126/science.1258479, 2015Spawning,doi:10.1126/science.aap9859, doi:10.1126/science.aar4005, Nasari:23, PhysRevLett.127.270602}, acoustics \cite{PhysRevX.6.021007, doi:10.1126/science.abd8872, PhysRevLett.121.085702, PhysRevLett.127.034301, PhysRevLett.118.174301, PhysRevApplied.16.014012}, and electrical circuits \cite{2020Generalized,0Observation, PhysRevB.107.085426, advs.202301128}. In NH systems, boundary effects no longer merely perturb the bulk states as in Hermitian systems, leading to the emergence of the non-Hermitian skin effect (NHSE) \cite{PhysRevLett.121.086803}, which refers to the phenomenon where an extensive number of eigenstates are exponentially localized at system's boundaries. This unique and widespread effect has sparked significant theoretical interest \cite{PhysRevLett.121.086803, PhysRevLett.124.086801, PhysRevLett.123.246801, PhysRevLett.125.226402, PhysRevX.14.021011, PhysRevB.109.165127,xiong2024nonhermitianskineffectarbitrary, PhysRevLett.131.116601,2021Universal,PhysRevLett.131.076401,HU202551}, and has been experimentally confirmed in various systems \cite{PhysRevLett.124.250402, PhysRevResearch.2.023265, PhysRevLett.125.206402, PhysRevA.107.L010202, PhysRevLett.129.086601, 2020Generalized,doi:10.1126/science.aaz8727,doi:10.1073/pnas.2010580117}.

Further investigations have classified NHSE into different orders \cite{PhysRevLett.128.223903, PhysRevLett.123.016805, PhysRevResearch.6.013213, PhysRevB.106.035425,2021Observation,PhysRevB.102.205118,PhysRevLett.131.116601}.
For a system with dimension $d$ and side length $L$, the $n$-th order NHSE refers to the presence of $\mathcal{O}(L^{d-n+1})$ skin modes scaling with the system size $L^{d}$.
While the occurrence of first-order skin effect (FOSE) in one-dimensional (1D) NH systems can be predicted by the spectral winding number of the energy spectrum under periodic boundary condition (PBC) \cite{2021Universal, PhysRevLett.125.126402}, a unified topological invariant to characterize these higher-order skin effects is still lacking.

In this Letter, we establish a unified non-Hermitian multipole theory to describe the skin effect in different orders. Specifically, we introduce a series of novel non-Hermitian multipole Bott indices (NHMBIs) defined in real space to characterize the non-Hermitian skin effect (NHSE) in different orders. For the first-order skin effect (FOSE) in 1D models, we show that the first-order NHMBI is strictly equivalent to the spectral winding number through analytical calculations. For systems exhibiting higher-order skin effects, we identify a richer correspondence between higher-order NHMBI and spectral geometry. In particular, we discover two distinct types of second-order skin effects (SOSEs). In the first type, we find that the non-zero region of second-order NHMBI on the complex energy plane exactly matches the area enclosed by paired boundary states under the periodic boundary condition (PBC) along one direction but the open boundary condition (OBC) along the other direction. In the second type, we show that the non-zero region of second-order NHMBI corresponds to the area swept by the boundary states under the generalized boundary condition (GBC) which interpolates PBC  and OBC spectra.

\textit{Non-Hermitian multipole Bott indices.}---{We start with the FOSE in 1D non-Hermitian systems with OBC. It has been proved that it is closely related to the spectral winding number $\nu$ of the Hamiltonian with PBC\cite{ PhysRevLett.125.126402}\begin{eqnarray}\label{eq1}
\nu=\frac{i}{2 \pi} \int_{-\pi}^{\pi} \frac{d \log \operatorname{det}\left[H(k)-E_{a} \right]}{d k} d k,
\end{eqnarray}
which measures the winding of the energy spectrum on the complex plane with respect to any complex reference energy $E_a$ as momentum $k$ transverses the Brillouin zone. The nonzero winding of the phase of $H(k)-E_{a}$ along the Brillouin zone, which indicates a nonvanishing persistent current of the system under PBC, leads to the presence of skin modes.}

{The NH topology of $H(k)$ can also be understood from the extended Hermitian Hamiltonian \cite{PhysRevX.8.031079} (See also Sec. I in Supplementary Materials (SM) \footnote{\label{fn}See Supplemental Material at http://link.aps.org/supplemental/xxx, for more details about the analytical derivation of the NHMBI, the explanation on the rationality of extending from first-order NHMBI to higher-order NHMBIs, and supplementary numerical results of NHSE in various 2D models, which include Refs.~\cite{PhysRevX.8.031079,PhysRevX.8.031079,doi:10.1126/science.aah6442,doi:10.1126/science.aah6442,PhysRevB.96.245115,PhysRevB.96.245115,PhysRevB.96.245115,PhysRevLett.80.1800,doi:10.1137/130920137,doi:10.1137/110852553,doi:10.1137/120885991,PhysRevB.56.8651,PhysRevLett.123.016805,PhysRevLett.132.136401,PhysRevLett.125.126402,WOS:000694666900037}.})
\begin{eqnarray}\label{eq2}
\tilde{H}(k,E_{a}):=\left(\begin{array}{cc}
0 & H(k)-E_{a} \\
H^{\dagger}(k)-E_{a}^{*} & 0
\end{array}\right), %.
\end{eqnarray}
which inherently respects a chiral symmetry $\Gamma \tilde{H}(k,E_{a})\Gamma^{-1}=-\tilde{H}(k,E_{a})$ with $\Gamma = \mathrm{diag}(1, -1)$. Such chiral-symmetric Hermitian systems belong to the AIII class in the topological classification table and are characterized by the winding number of $\tilde{H}(k,E_{a})$
\cite{PhysRevLett.113.046802,Ryu_2010,RevModPhys.88.035005}, which is equivalent to the NH spectral winding number $\nu$ characterizing the FOSE of $H(k)$.}

{Given that recent work demonstrates the real space representation of the winding number for 1D chiral-symmetric topological insulators \cite{PhysRevB.103.224208}, we derive
{the real-space first-order NHMBI for FOSE}
\begin{equation}
P_{x}\left( E_{a}\right)=\frac{1}{2 \pi i} \operatorname{Tr}\left[\log \left(\mathcal{P}^{A}_{x} \mathcal{P}^{B\dagger}_{x}\right)\right],\label{Px}
\end{equation}
with the dipole operator
\begin{equation}
\mathcal{P}_{x}^{S}=U_{S}^{\dagger}\left[ \sum_{R, \alpha } \exp\left({i \frac{2 \pi x}{L_{x}} }\right) |R, \alpha\rangle\langle R, \alpha|\right]U_{S}.
\end{equation}
Here $U_S$, for $S=A, B$, are unitary matrices obtained by the singular value decomposition (SVD) of $h=H-E_{a}$: $h=U_{A}\Sigma U_{B}^{\dagger}$ with $\Sigma$ being a diagonal matrix composed of singular values. $|R, \alpha\rangle$ represents the $\alpha$ orbital within the unit cell at $R$. We present the detailed derivation in SM \footnotemark[\value{footnote}].}

Furthermore, we generalize the idea to higher-order NHSE based on
the multipole moment theory used to describe higher-order topological insulators \cite{PhysRevLett.128.127601, lin2024probing, li2024exact, luo2024spin, doi:10.1126/science.aah6442, PhysRevB.96.245115}, we proposed the following NHMBIs in real space to describe the second and third-order NHSE,
\begin{eqnarray}
Q_{xy}\left( E_{a}\right)&=&\frac{1}{2 \pi i} \operatorname{Tr}\left[\log \left(\mathcal{Q}^{A}_{xy} \mathcal{Q}^{B\dagger}_{xy}\right)\right],\label{Qxy}\\
O_{xyz}\left( E_{a}\right)&=&\frac{1}{2 \pi i} \operatorname{Tr}\left[\log \left(\mathcal{O}^{A}_{xyz} \mathcal{O}^{B\dagger}_{xyz}\right)\right],\label{Oxyz}
\end{eqnarray}
where the multipole moment operators are
\begin{eqnarray}\label{S23}
\mathcal{Q}_{xy}^{S}=U_{S}^{\dagger}\left[ \sum_{R, \alpha } \exp\left({i \frac{2 \pi xy}{L_{x}L_{y}}}\right) |R, \alpha\rangle\langle R, \alpha|\right]U_{S},\\
\mathcal{O}_{xyz}^{S}=U_{S}^{\dagger}\left[ \sum_{R, \alpha } \exp\left({i \frac{2 \pi xyz}{L_{x}L_{y}L_{z}} }\right) |R, \alpha\rangle\langle R, \alpha|\right]U_{S}.
\end{eqnarray}
Further explanation for the rationality behind the generalization from \eqref{Px} to \eqref{Qxy} and \eqref{Oxyz} is provided in SM~\footnotemark[\value{footnote}]. Upon examining these formulas, it becomes evident that the calculation of the NHMBI only requires performing an SVD on the non-Hermitian Hamiltonian $h$. This indicates that the NHMBI can be employed to diagnose the occurrence of NHSE in any NH system, even if $h$ is non-diagonalizable.
Additionally, in order to avoid the complications associated with ill-defined multipole moment operators under PBC \cite{PhysRevB.100.245133}, we follow previous research \cite{li2024exactuniversalcharacterizationchiralsymmetric} and use OBC for the NHMBI calculation \footnotemark[\value{footnote}].

\textit{Skin corner weight.}---To qualitatively characterize the skin modes of NH systems under OBC, we define the corner weight as \cite{PhysRevLett.131.116601}:
\begin{eqnarray}
w_{\mathrm{CW}}^{(N)}\left(E_{n}\right) \equiv \sum_{\mathbf{r}} \sum_{i=1}^{N} (-1)^{i} \left|\psi_{n}^R(\mathbf{r})\right|^{4} e^{-\left|\mathbf{r}-\mathbf{r}_{i}\right| / \xi}, \label{w_CW}
\end{eqnarray}
where the sum runs over all lattice positions \(\mathbf{r}\), and \(\psi_{n}^R(\mathbf{r})\) is the right eigenfunction corresponding to the eigenvalue \(E_{n}\). The coordinates \(\mathbf{r}_{i}\) \((i = 1, \ldots, N)\) denote the positions of the \(i\)-th corner of an \(N\)-sided polygon, labeled sequentially in a clockwise direction starting from the bottom-left corner. In this work, we focus on 1D chains ($N=2$) and 2D square samples ($N=4$). The magnitude of \(w_{\mathrm{CW}}^{(N)} \left(E_{n}\right)\) quantifies the localization strength of skin modes. Note that $\xi$ should be chosen much smaller than the system size (i.e., $\xi\ll L$). The sign of \(w_{\mathrm{CW}}^{(N)} \left(E_{n}\right)\) indicates the localization direction e.g. for 1D systems, \(\mathbf{r}_i\) (\(i = 1, 2\)) represents two endpoints of a finite chain, and positive (negative) values of $w_{\mathrm{CW}}^{(2)}( E_{n})$ correspond to skin modes localized at the right (left) end of the system.

\begin{figure}
    \centering
    \includegraphics[width=1\columnwidth]{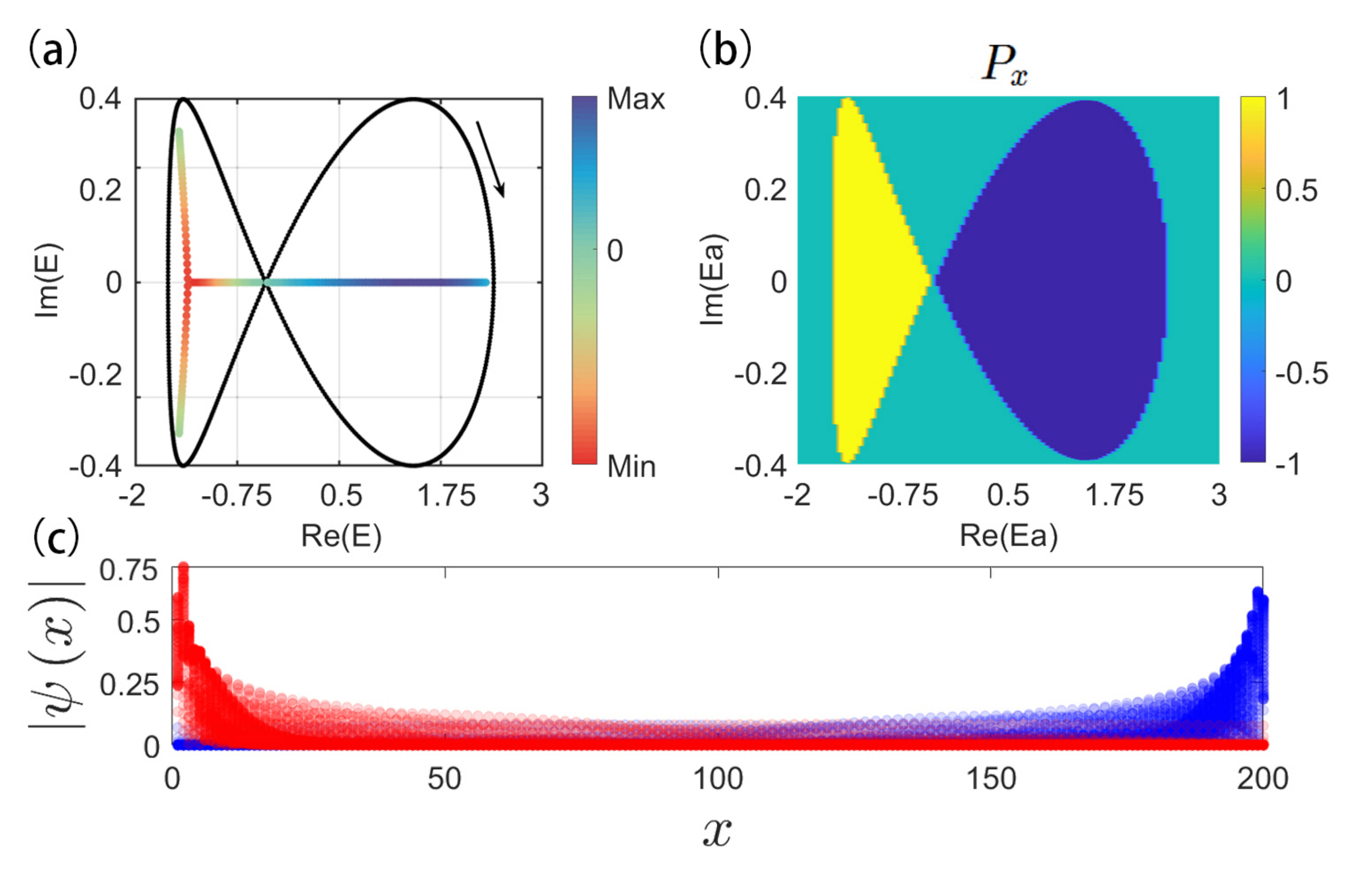}
\caption{\label{fig1:FOSE} (a) The OBC (color dots) and PBC (black dots) spectrum of the model in Eq.~(\ref{eq5}). The parameters used here are $t=1$, $t^\prime=0.4$, and the lattice size is {$L=200$}. The color of dots represent $w_{\mathrm{CW}}^{(2)}( E_{a})$ of the eigenstates, with parameters $r_{1}=1$, $r_{2}=L$, and $\xi=10$.
A positive (negative) value of \(w_{\mathrm{CW}}^{(2)}( E_{n})\) indicates localization of the SE mode at the right (left) side of the system. Black arrow represents the winding direction of the PBC energy spectrum. (b) Distribution of $P_x(E_{a})$ in the complex reference energy plane. (c) Spatial profiles of all normalized eigenfunctions under OBC. }
\end{figure}

\begin{figure*}%[!htbp]
    \centering
    \includegraphics[width=1\textwidth]{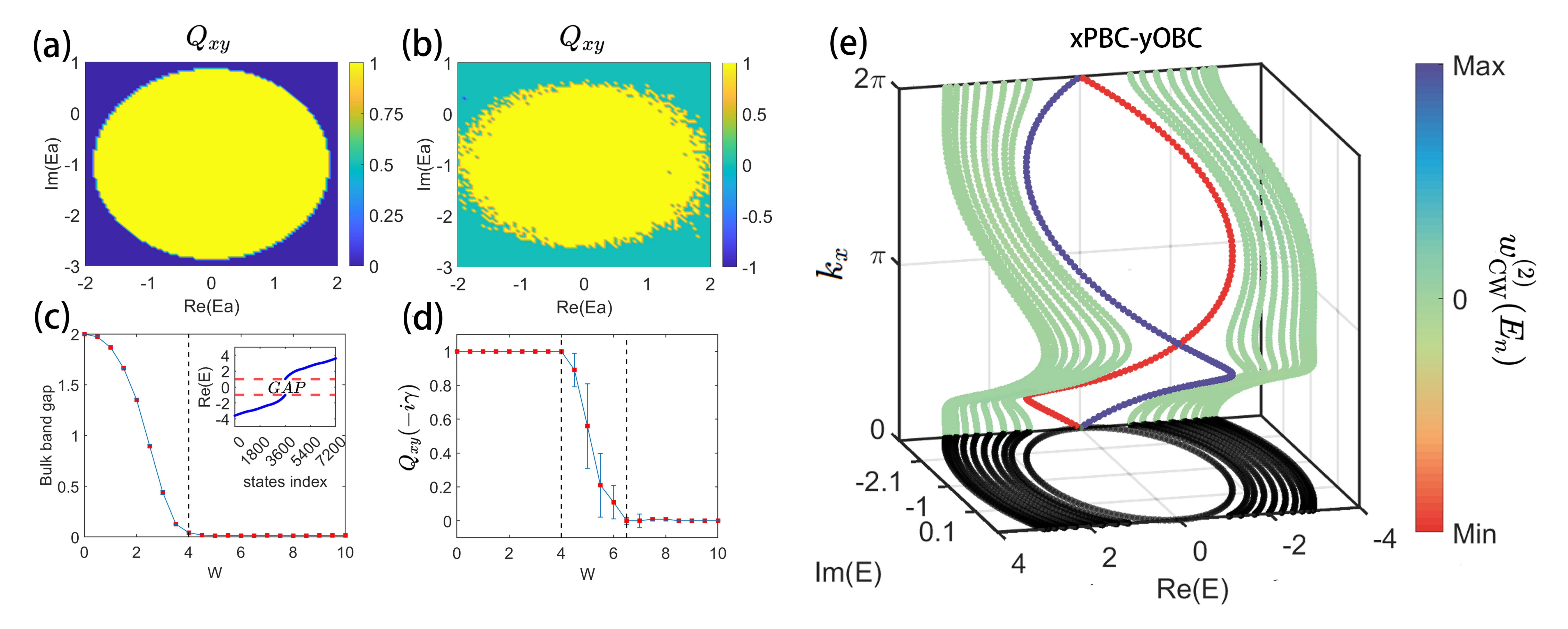}
\caption{\label{fig:sose} Distribution of $Q_{xy}(E_{a})$ on the complex reference energy plane for the system (a) without disorder ($W = 0$) and (b) with a finite disorder of $W=2$, respectively. (c) The gap of the real part of energy Re($E_n$) as a function of disorder strength $W$ in the PBC spectrum. Inset represents Re$(E_n)$ in ascending order at $W=0$. (d) Evolution of $Q_{xy}\left(-i\gamma \right)$ with $W$ in 2D square disordered systems with size $60\times 60$. For each $W$, the configuration average is performed over 100 realizations. Two black dashed lines represent $W=4$ and $W=6.5$, respectively. (e) $x$PBC-$y$OBC spectrum for the strip system with $L_{y}=60$. The color of dots in the spectrum represents the corresponding {corner weight} $w_{\mathrm{CW}}^{(2)} \left( E_{n}\right)$ of the eigenstate, with parameters $\xi=1$. Black dots at the bottom are colored dots' projections on the complex energy plane. It can be seen that the circle formed by the TS is consistent with the circular boundary of the non-zero region in (a). The parameters $\gamma=1$ and $\lambda=2$ are used for all calculations.
}
\end{figure*}

\textit{FOSE in 1D systems.}---
To verify the effectiveness of the first-order NHMBI, we first consider a 1D single-band model that exhibits the FOSE \cite{PhysRevLett.125.126402, PhysRevB.109.035119},
\begin{eqnarray}\label{eq5}
H_{\mathrm{1D}}(k)=te^{-i k}+te^{i k}+t^\prime e^{i2k},
\end{eqnarray}
where $t$ and $t^\prime$ are real parameters. The FOSE in this model can be captured by the first-order NHMBI $P_x(E_{a})$ in Eq.~\eqref{Px}.
Fig.~\ref{fig1:FOSE}(a) shows the energy spectrum of the 1D NH model (Eq.~\eqref{eq5}) under OBC and PBC, respectively. One observes that the PBC spectrum forms an asymmetric $\infty$-shape loop that has opposite spectral winding numbers ($\nu=\pm 1$) with respect to the reference energy $E_a$ inside the left and right enclosed areas,
which signifies the existence of NHSE\cite{PhysRevLett.125.126402}. By analyzing $w_{\mathrm{CW}}^{(2)}\left(E_{n}\right)$ of the OBC spectrum, we find that all eigenstates inside the left (right) area with $\nu=1$ ($\nu=-1$) are skin modes {spatically} localized on the left (right) side of the chain [see Fig.~\ref{fig1:FOSE}(c)]. Consequently, an extensive number [i.e., $\mathcal{O}(L)$] of skin modes appear in the 1D NH model with length $L$ under OBC. For comparison, we plot $P_x(E_{a})$ as shown in Fig.~\ref{fig1:FOSE}(b). Remarkably, the left and right areas exhibiting $P_x(E_{a})=+1$ and $-1$ {in the complex energy plane} are exactly the same as that with $\nu=\pm 1$ in Fig.~\ref{fig1:FOSE}(a). This indicates that $P_x(E_{a})$ is consistent with the spectral winding number $\nu$ in the complex reference energy plane and, therefore, corresponds to the presence of skin modes under OBC.

\textit{Type-I SOSE and disorder-induced transition.}---{Next, we verify the applicability of the real-space NHMBI for higher-order skin effects. As an example, we consider a 2D prototypical NH model for SOSE on a square lattice with two sublattices \cite{PhysRevB.102.205118},
\begin{eqnarray}\label{eq7}
{H}_{\mathrm{2D}}(\kk)&=&-i\left( \gamma + \lambda \cos k_{x}\right) \sigma_{0}-\lambda\sin k_{x} \sigma_{z}\nonumber\\
& &+\left(\gamma+ \lambda \cos k_{y}\right) \sigma_{y}+\lambda \sin k_{y}  \sigma_{x},\label{H_2D}
 \end{eqnarray}
where $\gamma$ and $\lambda$ are real parameters. $\left\{\sigma_{x}, \sigma_{y}, \sigma_{z}\right\}$ and $\sigma_{0}$ are Pauli matrices and identity in sublattice space, respectively. %(1)
First, we find that $(P_x, P_y)=\bm{0}$ in the whole reference energy plane,
indicating the absence of FOSE in Eq.~\eqref{H_2D} along the $x$ and $y$ directions. Here we only focus on rectangle geometry and did not take the geometry-dependent skin effect into consideration \cite{2021Universal}. Nevertheless, in a rectangular geometry with OBC in both $x$ and $y$ directions, $\mathcal{O}(L)$ skin modes appear at the corners for $|\gamma/\lambda|<1$, signaling the occurrence of SOSE. We calculate $Q_{xy}$ in Eq.~\eqref{Qxy} for the model. As shown in Fig.~\ref{fig:sose}(a), a circular area centered at $E^0_a = -i\gamma$ exhibits nonzero $Q_{xy}=1$ in the complex energy plane. The perimeter of the circle is captured by $E_{a}(\theta)=-i \gamma-i \lambda e^{-i \theta},\; \theta \in[0,2 \pi)$. In contrast, $Q_{xy}$ vanishes in the whole complex energy plane for $|\gamma/\lambda| >1$, which is compatible with the absence of SOSE. Previous studies \cite{PhysRevB.102.205118, PhysRevLett.131.116601, PhysRevB.104.L121101} have shown that the bulk energy spectrum is separated by a so-called line gap which closes only at $|\gamma/\lambda|=1$, accompanies a topological phase transition of SOSE.
Therefore, the sudden evaporation of $Q_{xy}$ accurately reflects the topological phase transition of SOSE.}

{The occurrence of SOSE in the model can also be understood from the spectrum under PBC along the $x$ direction but OBC along the $y$ direction (i.e., $x$PBC-$y$OBC). {As shown in the bottom black projections of Fig.~\ref{fig:sose}(e), $\mathcal{O}(L)$ modes form a circle at the center of the $x$PBC-$y$OBC spectrum. These modes are spatially confined to both open edges, identifying them as localized edge modes. Interestingly, we noticed that the circular region enclosed by these localized edge modes in Fig.~\ref{fig:sose}(e) precisely coincides with the nonzero region of $Q_{xy}$ in Fig.~\ref{fig:sose}(a).}
Furthermore, by tracing the evolution of these localized edge modes while varying $k_x$ from 0 to $2\pi$ [see Fig.~\ref{fig:sose}(e)], we found that the edge modes on top and bottom edges (denoted by red and blue dots {according to their corner weight $w_{\mathrm{CW}}^{(4)}\left( E_{n}\right)$}) wind around clockwise and anticlockwise, respectively, giving rise to opposite spectral winding numbers $\nu=\pm 1$. Conversely, there is no winding in the bulk modes (green dots). Therefore, the localized edge mode can be regarded as two effective 1D NH models with opposite $\nu$ at the top and bottom edges. When OBC are imposed in both directions, the edge modes evolve into $\mathcal{O}(L)$ skin endpoint modes at opposite ends of two edges, which are nothing but the skin corner modes at opposite corners, leading to the SOSE. Given that the FOSE corresponds to the persistent charge current under PBC \cite{2021Universal, PhysRevLett.125.126402}, we therefore, propose that the SOSE corresponds to the persistent dipole current localized on opposite edges under the hybrid $x$PBC-$y$OBC, as illustrated in Fig.~S6 in SM~\footnotemark[\value{footnote}].
}

In addition to translational invariant NH systems, our proposed NHMBI, defined in real space, can also diagnose the disorder-driven transition of SOSE. Specifically, we introduce a disorder potential in Eq.~\eqref{H_2D} through random on-site energy with a uniform distribution within the interval $[-W, W]$. Compared to the clean limit ($W=0$), it is evident that the nontrivial region with $Q_{xy}(E_{a})=1$ shrinks in the presence of disorder ($W=2$), and the ragged boundary {of the nontrivial region} becomes less well-defined, as shown in Fig.~\ref{fig:sose}(a,b).
To illustrate the disorder-induced phase transition, we plot the bulk energy gap $E_g^\mathrm{Re}$ for the real part of eigenvalues $\mathrm{Re}(E_n)$ and $Q_{xy}$ at $E_a=-i\gamma$ [i.e., $Q_{xy}\left( -i \gamma \right)$] as functions of the disorder strength $W$. As shown in Fig.~\ref{fig:sose}(c,d), $E_g^\mathrm{Re}$ gradually decreases with increasing disorder strength $W$, eventually closing at $W=4$. However, $Q_{xy}\left( -i \gamma \right)$ remains quantized at 1 until $W > 4$, after which it begins to decrease with significant fluctuations reflecting different disorder configurations with $Q_{xy}\left( -i \gamma \right) =0$ or 1. When $W > 6.5$, $Q_{xy}\left( -i \gamma \right)$ vanishes to zero, signaling that the system has transitioned into a trivial phase without SOSE.

\begin{figure}
    \centering
    \includegraphics[width=1.0\linewidth]{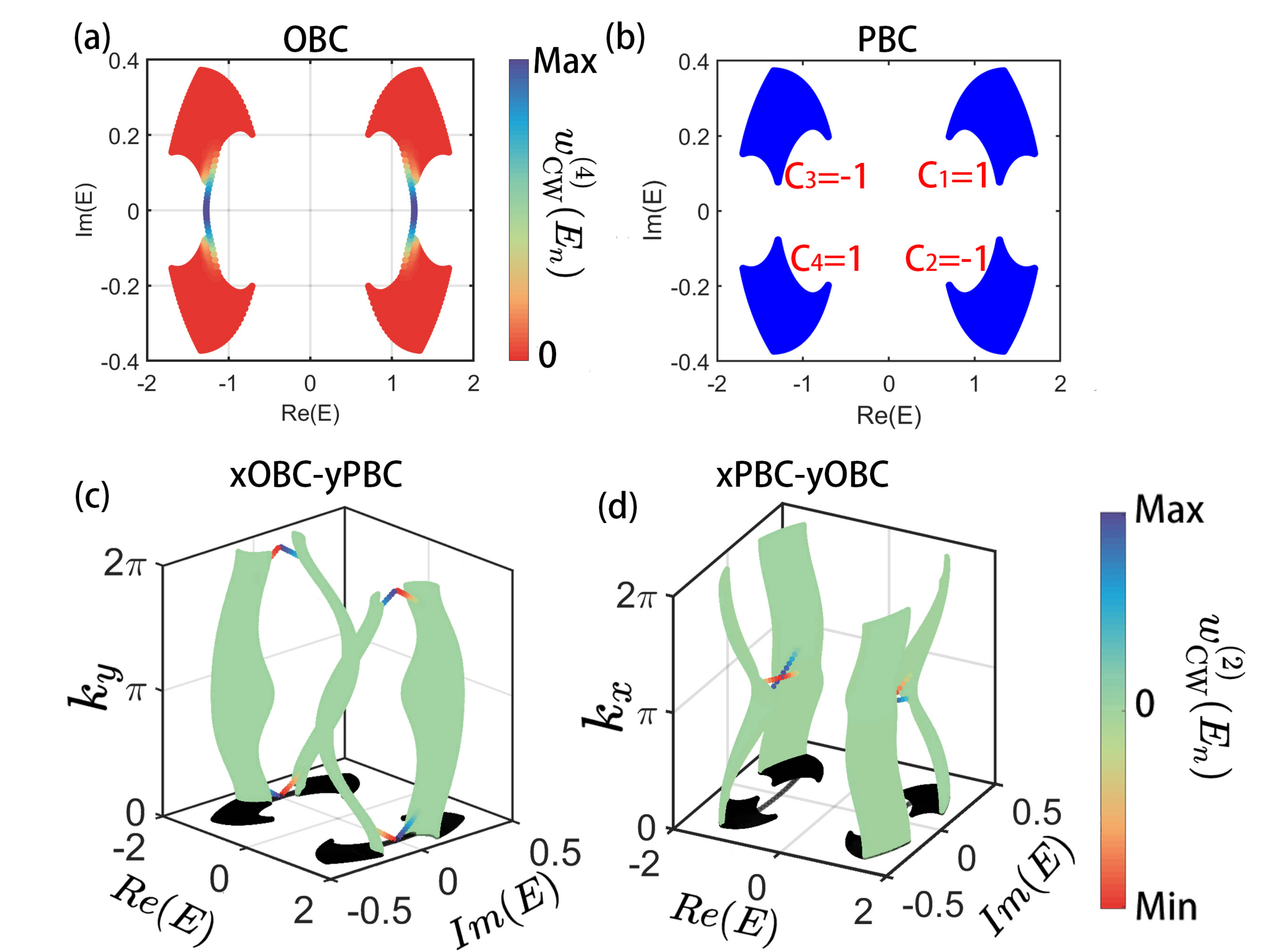}
    \caption{The energy spectrum of the type-II SOSE model under various boundary conditions with parameters $\left(t, t_{x}, t_{y}, \delta, \delta^{\prime}\right)=(0.3,1,1,0.4,-0.6)$
 (a) The OBC energy spectrum. The color of each eigen energy corresponds to {corner weight} $w_{\mathrm{CW}}^{(4)}\left( E_{n}\right)$ of the eigenstate with $\xi=1$. (b) Energy spectrum under PBC. The non-Hermitian Chern numbers corresponding to different occupancy states are highlighted in red within the graph. (c,d) The spectrum under (c) $x$OBC-$y$PBC and (d) $x$PBC-$y$OBC. The color of each eigen energy corresponds to {corner weight} $w_{\mathrm{CW}}^{(2)}\left( E_{n}\right)$ of the eigenstate with $\xi=1$.}
    \label{fig3:enter-label}
\end{figure}

\begin{figure}%[!htbp]
    \centering
    \includegraphics[width=1.0\linewidth]{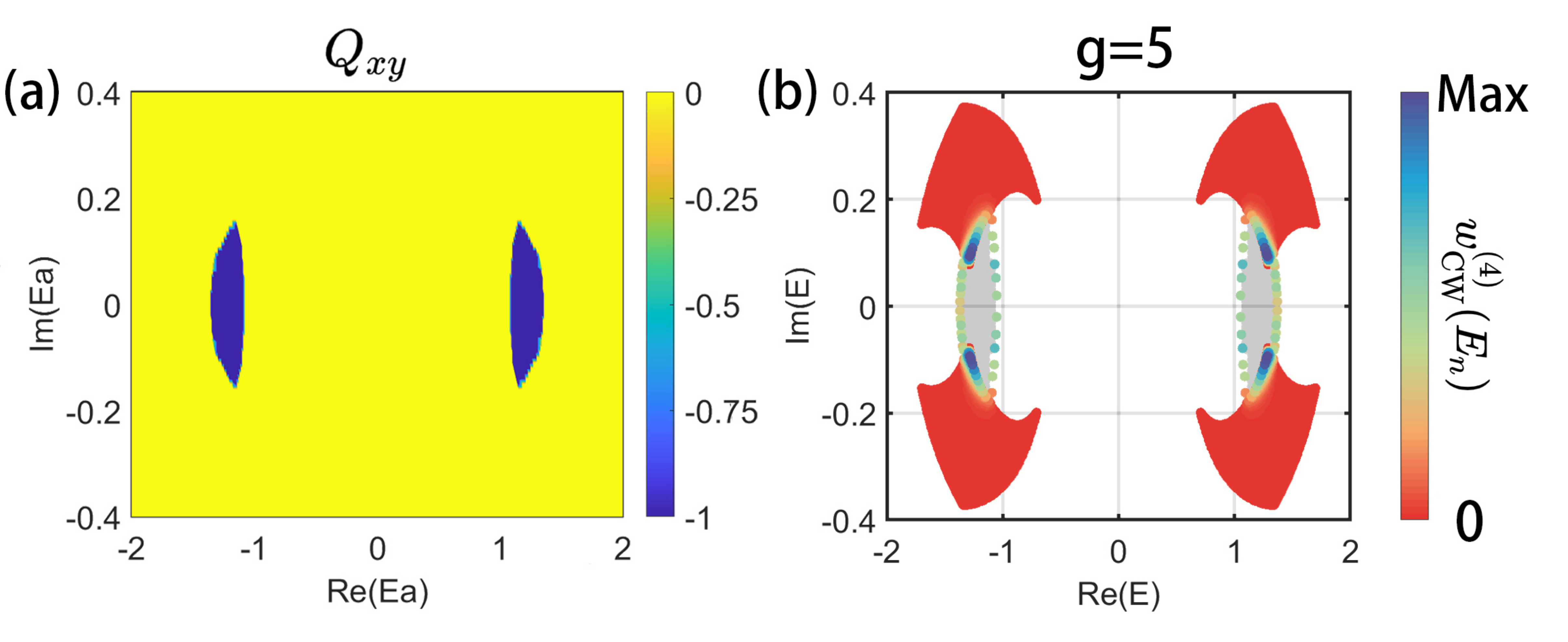}
    \caption{
    (a) The distribution of $Q_{xy}\left( E_{a}\right)$
  across the complex reference energy plane for the type-II SOSE model system. (b) The energy spectrum of the system with GBC at $g=5$, where the shaded region denotes the non-zero region of $Q_{xy}\left( E_{a}\right)$. Notably, the area enclosed by the boundary states in this scenario aligns with the non-zero region of $Q_{xy}\left( E_{a}\right)$. The color of each eigen energy corresponds to {corner weight}  $w_{\mathrm{CW}}^{(4)}\left( E_{n}\right)$ of the eigenstate with $\xi=1$. (a) and (b) share the same model parameters as depicted in Fig.~\ref{fig3:enter-label}.}
    \label{fig4:enter-label}
\end{figure}

\textit{Type-II SOSE and generalized boundary conditions}.---In addition to the above model of SOSE (dubbed type-I), we also notice an anomalous case where SOSE emerges even though the boundary modes under partial-PBC, i.e., $\alpha$PBC-$\beta$OBC ($\alpha,\beta=x,y$), fail to wind around and enclose finite areas
\cite{PhysRevLett.123.016805}, which is referred as type-II SOSE.

{Fundamentally, the distinction between type-I and type-II SOSE models stems from their different topological characteristics. Type-I SOSE system is structurally analogous to a weak topological insulator consisting of an array of 1D topological chains that are non-reciprocally coupled \cite{PhysRevLett.98.106803}. In a square geometry, topological modes originally localized on opposing ends of the coupled chains are squeezed to corners due to the nonreciprocal coupling at the boundaries, giving rise to SOSE. }

{In contrast, the type-II SOSE system follows a different topological paradigm. Here, topologically localized modes, which initially appear along all open edges, accumulate at the corners due to boundary-induced local nonreciprocity, ultimately leading to SOSE. Since type-II SOSE arise from simultaneous higher-order interplay between topological localization and nonreciprocal boundary pumping, it cannot be reduced to conventional 1D skin effect descriptions.} Despite the lack of spectral winding in this scenario, we show that the NHMBI remains a valid quantity for characterizing the emergence of type-II SOSE.

Specifically, we consider the 2D NH model exhibiting type-II SOSE, as presented in Ref.~\cite{PhysRevLett.123.016805}. The Hamiltonian for this system is given by
\begin{eqnarray}\label{type-IISOSE}
H_{\mathrm{2D}}(\boldsymbol{k}) &=& \left(t_{x} + t \cos k_{x}\right) \tau_{0} \otimes \sigma_{x} + i \delta \tau_{z} \otimes \sigma_{y} \nonumber \\
&+& \left(t_{y} + t \cos k_{y}\right) \tau_{x} \otimes \sigma_{z} - i \delta^{\prime} \tau_{y} \otimes \sigma_{0} \nonumber\\
&+& t \sin k_{x} \tau_{0} \otimes \sigma_{y}-t \sin k_{y} \tau_{y} \otimes \sigma_{z},
\end{eqnarray}
where \(t_x\), \(t_y\), \(t\), \(\delta\), and \(\delta'\) are real parameters. This model can be viewed as a grid of two non-reciprocal SSH models in the \(x\)- and \(y\)-directions. For the parameter set \(\left(t, t_x, t_y, \delta, \delta'\right) = (0.3, 1, 1, 0.4, -0.6)\), the system exhibits skin corner modes under OBC, as shown in Fig.~\ref{fig3:enter-label}(a).
This arises from the boundary skin effect on the first-order topological modes dictated by nontrivial Chern numbers [Fig.~\ref{fig3:enter-label}(b)]. Specifically, the breaking of reciprocity at the boundary induces the nonreciprocal pumping which leads to boundary skin modes accumulation along the OBC direction, resulting in SOSE.
We further calculate the energy spectrum under partial-PBC, as shown in Fig.~\ref{fig3:enter-label}. {Unlike the SOSE studied above, the boundary skin modes of Eq.~\eqref{type-IISOSE} do not form closed loops with nonzero spectral winding numbers in the energy spectrum under either $x$PBC-$y$OBC or $x$OBC-$y$OBC conditions}.

Surprisingly, upon calculating \(Q_{xy}(E_a)\), we observe non-zero regions on the complex energy plane [see Fig.~\ref{fig4:enter-label}(a)], {while $(P_x, P_y)=\bm{0}$ across the entire reference energy plane}, which is consistent with the occurrence of SOSE. {To address the bulk-boundary correspondence conundrum}, we introduce a Hamiltonian under generalized boundary conditions (GBC) given by
$H_\mathrm{GBC}\left( g\right)\equiv H_\mathrm{OBC}+e^{-g} H_{1 \leftrightarrow N_{x}}+e^{-g} H_{1 \leftrightarrow N_{y}}$,
where $H_{1 \leftrightarrow N_{\alpha}}$ denotes the hoppings between the first and last unit cells along the $\alpha$-direction $\left(\alpha=x,y \right)$, and \(g\) is a real parameter \cite{WOS:000694666900037, lei2024activatingnonhermitianskinmodes}. When \(g = 0\), it corresponds to PBC, while as \(g \to \infty\), it approaches OBC.
By applying GBC, we observe that as \(g\) increases, the boundary skin modes initially localize at the edges and gradually accumulate toward the corners (see SM~\footnotemark[\value{footnote}]).  Interestingly, we find that the boundary skin modes enclose the largest area for \(g = 5\) {in the energy spectrum}, as shown in Fig.~\ref{fig4:enter-label}(b). As \(g\) continues to increase, this area diminishes and the skin spectrum converges to the OBC limit [see Fig.~\ref{fig3:enter-label}(a)]. Notably, this maximum area swept by boundary skin modes corresponds exactly to the non-zero region of \(Q_{xy}(E_a)\) {in the complex energy plane} [as presented in Fig.~\ref{fig4:enter-label}(a)], indicating a direct connection between the geometric properties of the GBC spectrum and the NHMBI in the type-II SOSE model.

\iffalse
{To quantitatively describe the relationship between GBC spectral evolution and NHMBI, we analyze the quantity under GBC: $I_\mathrm{GBC}\left( E_{a}\right) = \sum _{n}\left|1/({E_{n}\left( g\right) -E_{a}})\right|$, which represents the absolute sum of the inverse energy gaps between the evolving spectrum $E_{n}\left( g\right)$ and a reference energy $E_{a}$ \cite{WOS:000694666900037}. This quantity serves as a diagnostic tool to track how many times $E_{n}\left( g\right)$ approaches $E_{a}$ as $g$ is varied. For 1D NH systems, the number of singular points of $I_\mathrm{GBC}\left( E_{a}\right)$ corresponds directly to the spectral winding number with $E_{a}$ as the reference point \cite{WOS:000694666900037}. The result of $I_\mathrm{GBC}\left( E_{a}\right)$ for the 2D type-II SOSE model \eqref{type-IISOSE} is presented in SM~\footnotemark[\value{footnote}]. When $E_a$ is chosen within the non-zero region of $Q_{xy}$, a singular point emerges in $I_\mathrm{GBC}\left( E_{a}\right)$ as $g$ increases from 0 to $\infty$. In contrast, selecting $E_a$ in the region where $Q_{xy}$ vanishes eliminates the singular point. This behavior mirrors the evolutionary pattern of $I_\mathrm{GBC}\left( E_{a}\right)$ observed in the 1D NH models under GBC. Thus, these findings further support our conclusion that the area swept by the boundary state aligns with the non-zero region of $Q_{xy}$, reinforcing the topological connection between the spectral evolution and NHMBI.}
\fi

\textit{Summary.}---We have introduced a series of NHMBI defined in real space to characterize NHSE in different orders.  For 1D NH systems, we rigorously prove that the first-order NHMBI is equivalent to the spectral winding number proposed in previous studies. This equivalence implies that, for 1D NH models, the area enclosed by the energy spectrum under PBC corresponds to the non-zero region of first-order NHMBI in the complex reference energy plane, which can be interpreted as a generalized bulk-edge correspondence.

For higher-order skin effects, this correspondence becomes even more intricate. We have identified two types of correspondences for SOSE. For type-I SOSE, the positive and negative winding formed by paired boundary states under partial-PBCs matches the non-zero region of second-order NHMBI. For type-II SOSE, the boundary states under partial-PBCs no longer form spectral winding, yet the second-order NHMBI remains effective in characterizing the SOSE. Interestingly, we found that by introducing the GBC to continuously deform the system from PBC to OBC, the area swept by the boundary states aligns with the non-zero region of the second-order NHMBI in the complex energy plane.

Furthermore, there are still several open questions that we believe are meaningful for further exploration. First, it remains to be seen whether other spectral geometries exhibit corresponding relationships with the non-zero region NHMBI or if a more unified connection can be established between spectral geometry and the NHMBI. Second, for NH systems with special symmetries (such as the $Z_{2}$ skin effect \cite{PhysRevLett.124.086801,guo2023anomalousnonhermitianskineffect}), it would be interesting to investigate whether NHMBI can be generalized to capture these phenomena. Finally, a compelling direction would be to explore whether the NHMBI proposed here can be directly related to measurable observables in experiment settings.

\begin{acknowledgments}
This work is supported by the National Key R\&D Program of China (Grant No. 2021YFA1401600), the National Natural Science Foundation of China (Grant No. 12474056), and the 2022 basic discipline top-notch students training program 2.0 research project (Grant No. 20222005).
The work was carried out at the National Supercomputer Center in Tianjin, and the calculations were performed on Tianhe new generation supercomputer. The high-performance computing platform of Peking University supported the computational resources.
\end{acknowledgments}


\end{document}